# Quantitative clarification of key questions about COVID-19 epidemiology

Yinon M. Bar-On[1*], Ron Sender[1*], Avi I. Flamholz[2], Rob Phillips[3,4], Ron Milo[1]
[1]Weizmann Institute of Science, Rehovot 7610001, Israel
[2]University of California, Berkeley, CA 94720, USA
[3]California Institute of Technology, Pasadena, CA 91125, USA
[4]Chan Zuckerberg Biohub, 499 Illinois Street, SF CA 94158, USA
* Equal contribution

Abstract:
**Modeling the spread of COVID-19 is crucial for informing public health policy. All models for COVID-19 epidemiology rely on parameters describing the dynamics of the infection process. The meanings of epidemiological parameters like $R_0$, $R_t$, the "serial interval" and "generation interval" can be challenging to understand, especially as these and other parameters are conceptually overlapping and sometimes confusingly named. Moreover, the procedures used to estimate these parameters make various assumptions and use different mathematical approaches that should be understood and accounted for when relying on parameter values and reporting them to the public. Here, we offer several insights regarding the derivation of commonly-reported epidemiological parameters, and describe how mitigation measures like lockdown are expected to affect their values. We aim to present these quantitative relationships in a manner that is accessible to the widest audience possible. We hope that better communicating the intricacies of epidemiological models will improve our collective understanding of their strengths and weaknesses, and will help avoid possible pitfalls when using them.**

**Introduction**

Quantitative characterization of the COVID-19 pandemic is a necessary precursor for efforts to model the spread of the virus. Epidemiological models rely on numerical parameters like the "generation interval" and "household attack rate" whose values and distributions are estimated from data about the outbreak. Such estimates are routinely reported in news media and used in epidemiological models that inform policy decisions, but the estimates are always uncertain, affected by the underlying data and assumptions, and the parameters themselves are mutually interrelated and often confusingly described. We previously compiled quantitative data about the basic properties of the SARS-CoV-2 virus and its interaction with the human body [(Bar-On et al. 2020)](Bar-On et al. 2020) to provide a "one-stop" source for clear explanations and annotated references about quantitative properties associated with this virus. Here, we aim to provide a similar resource focused on numbers used in epidemiological models designed to forecast the spread of the virus. In an accompanying study, we provide a compendium of such epidemiological parameters and their values. Here, we describe insights gleaned from reading the epidemiological literature and collecting estimates for each parameter.

We offer these insights in the form of short vignettes that demonstrate how a quantitative description of the COVID-19 epidemic can help us understand information we receive from medical professionals and news outlets, and improve our decision making on the personal, institutional, and governmental levels. For example, if a person receives a positive PCR test, what are the odds she actually has COVID-19? How are these odds affected by the structure of a local, regional or corporate testing program? Several vignettes also aim to highlight possible biases that arise when estimating key epidemiological parameters. For example, we describe how underlying assumptions can greatly affect estimates of the basic and effective reproduction numbers (variously called $R_0$, and $R_e$ or $R_t$), a fundamental and widely-reported epidemiological parameter. Finally, several vignettes aim to help non-epidemiologists become acquainted with key parameters that are less well-known than $R_0$ but no less important, for example the generation interval or the dispersion of the offspring distribution.

Our collection of vignettes is by no means comprehensive and does not replace detailed analysis of the kind that is performed in dedicated studies. In addition, several key parameters remain poorly characterized such as the infectiousness of asymptomatic cases, the infectious dose of SARS-CoV-2 or the effect of climate on the transmission of COVID-19. Once relevant data regarding these parameters will amass, future analysis should address them as well.

**Definitions**

The basic reproduction number - $R_0$: The basic reproduction number of an infection, $R_0$, is the expected number of secondary infections generated by an average infectious case in an entirely susceptible population.

The effective reproduction number - $R_e$ or $R_t$: The effective reproduction number, $R_t$ (also denoted by R(t), R and $R_e$) is the number of secondary cases generated by an infectious case at a given moment (t) once an epidemic is underway. Unlike the basic reproductive number, $R_t$ is time and situation specific, used to characterize pathogen transmissibility during an epidemic, and enabling assessment of the effectiveness of interventions.

Generation interval: The average time between infection events in an infector-infectee pair.

Serial interval: The average time between symptom onsets in an infector-infectee pair.

The dispersion of the offspring distribution - k: The "offspring distribution" of an epidemic is the distribution of the number of secondary cases due to each primary infection. The mean of this distribution is $R_t$ by definition. The offspring distribution is generally considered to adopt a negative binomial distribution which is characterized by a mean ($R_t$) and a dispersion parameter k. Smaller values of k indicate greater heterogeneity. In general k<<1 is associated with a high number of 'superspreaders' - individuals that infect much more than the average $R_t$. Values of k>1 are associated with a low number of 'superspreaders', as the number of secondary cases of each primary infection is relatively even.

Household secondary attack rate: The fraction of the infectee's household members that they infect on average.

**Table of Contents**

## How does the basic reproduction number, $R_0$, relate to the doubling time of the epidemic?

The basic reproduction number of an infection, $R_0$, is defined as the expected number of secondary infections generated by a single infected person when the population is still close to entirely susceptible (i.e. at the beginning of an epidemic, when only a small fraction of the population has been infected). $R_0$ describes the number of secondary infections, but gives no information about the timing of secondary infections. The characteristic time between when a primary and secondary infection is called the "generation interval". Both $R_0$ and the generation interval vary between individuals, but we can define average values for both of them to understand how they govern the spread of SARS-CoV-2. This basic model assumes that the pandemic starts with a single infection and the total number of infections, N(t), grows exponentially over time so that $N(t) = 1 \times R_0^{(t/g)}$. Here *t* is the time in days and *g* is the generation interval, also in days, so that *t/g* is unitless. Given this model, the total number of cases should increase by a factor of $R_0$ every *g* days.

Let's consider an example where the number of cases doubles every 4 days. If the generation interval *g* is known to be ≈4 days as well, then we infer that $R_0 \approx 2$ (as $N(t+4)/N(t) = 2 = R_0^{(4/4)} = R_0$). If instead the generation interval is known to be 8 days, then a doubling time of two days implies $R_0 \approx 4$ (as $N(t+4)/N(t) = 2 = R_0^{(4/8)}$). As such, it's important to have a high-quality estimate of the generation interval, *g*, when attempting to make inferences about the viral reproduction number $R_0$. When there is sufficient data to estimate the distribution of generation intervals (rather than just the average value), this simple connection between $R_0$ and the doubling time ($\tau$) can be still be applied. See (J. Wallinga and Lipsitch 2007) for detailed discussion.

Early in the COVID-19 pandemic, there was substantial uncertainty about the distribution of generation intervals and various researchers used different values to derive estimates of $R_0$. As such, one should take care in comparing $R_0$ values between studies, making sure to evaluate the assumptions about the doubling time and generation intervals that were used. Figure 1 describes the relationship between reported values of $R_0$, doubling times and serial intervals from various studies of the transmission at the beginning of the epidemic in Wuhan, China. The serial interval (the interval between infector and infectee symptom onset) is sometimes used as a proxy for the generation interval because it is simpler to estimate from patient reports and contact tracing data. $R_0$ estimates depend on the assumed serial interval and doubling time in a complex fashion, leading to wide variation in reported values ranging from $R_0 \approx$ 2-6.

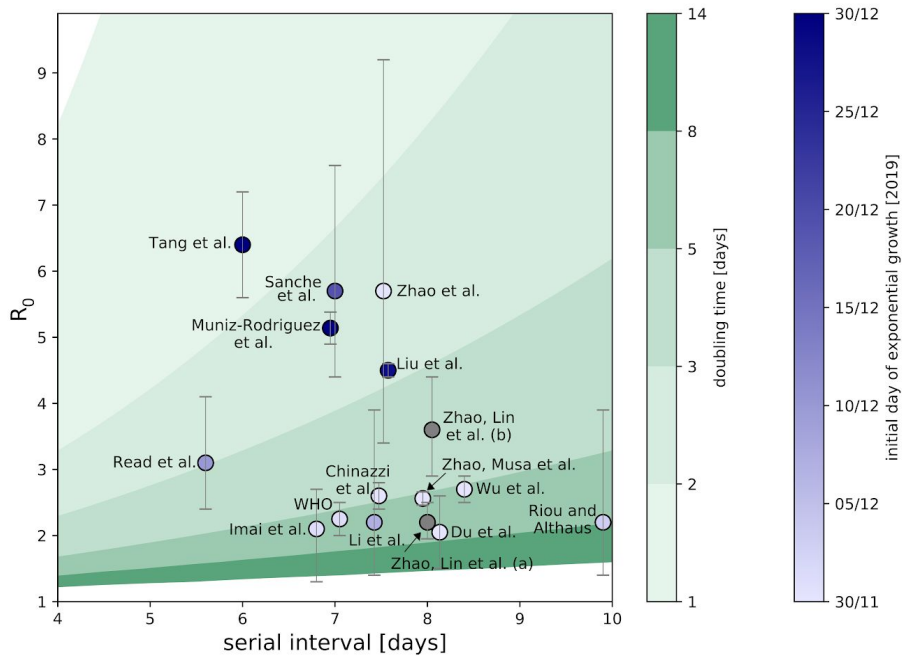

**Figure 1: The inter-relation between the estimates of $R_0$, the doubling time and the serial interval.** The estimates are presented as a scatter plot versus the assumed mean serial interval (equal to the mean generation interval). In a few cases where the assumption regarding the serial interval wasn't explicit, we extracted it from other assumptions of the model. The background of the figure is colored by estimates of the doubling times, derived from $R_0$, the mean serial interval and an assumption that the serial interval is gamma distributed with a coefficient of variation of 50% (similar to explicit assumptions in some of the sources). In the few cases in which the source provides its estimate for the doubling time, the results are similar (but not identical). Markers of studies with identical serial intervals were slightly displaced to avoid overlap. The markers are colored by the date on which the model assumes that exponential growth began, except for (S. Zhao et al. 2020) which did not report a start date.

**What causes large discrepancies between estimates of $R_0$ for the same place?**
Although sources differ in their assumptions about the distribution of generation intervals, these differences are not the largest factor affecting the estimated $R_0$. We find that a larger influence arises from the presumed date on which the epidemic started, denoted $t_0$. Sources that assumed a start date near Dec. 1 2019 considered a much longer time period than studies assuming a start date in January 2020. Fitting over longer time periods resulted in slower growth apparent rates in general, as can be seen from the following example. Let's assume that on Jan. 20th there were 1024 cases. To get to 1024 cases from 1 patient, it would take 10 doublings ($2^{10}$ = 1024). So if exponential growth began on the 31st of December that would mean a doubling time of 20/10 = 2 days. But if the exponential growth was assumed to begin on December 1st, that would imply a doubling time of 50/10=5 days. This effect can be seen in Figure 1, where the color indicates the $t_0$ assumed in each source. Figure 2 shows the correlation between the assumed $t_0$ and the inferred $R_0$. Linear regression gives $R^2$=0.52, indicating that $t_0$ is a relatively strong predictor of the inferred $R_0$ value. From the analysis of the difference sources, it appears that only one source directly inferred the initial date of exponential growth from the model (Steven Sanche et al. 2020) while the others assumed that growth began on a specific date, even though results are sensitive to this assumption. Therefore, it seems that wide variation in the exponential growth rate and reproduction number across studies of the Wuhan outbreak stem from variation in the choice of a "start date" for the epidemic.

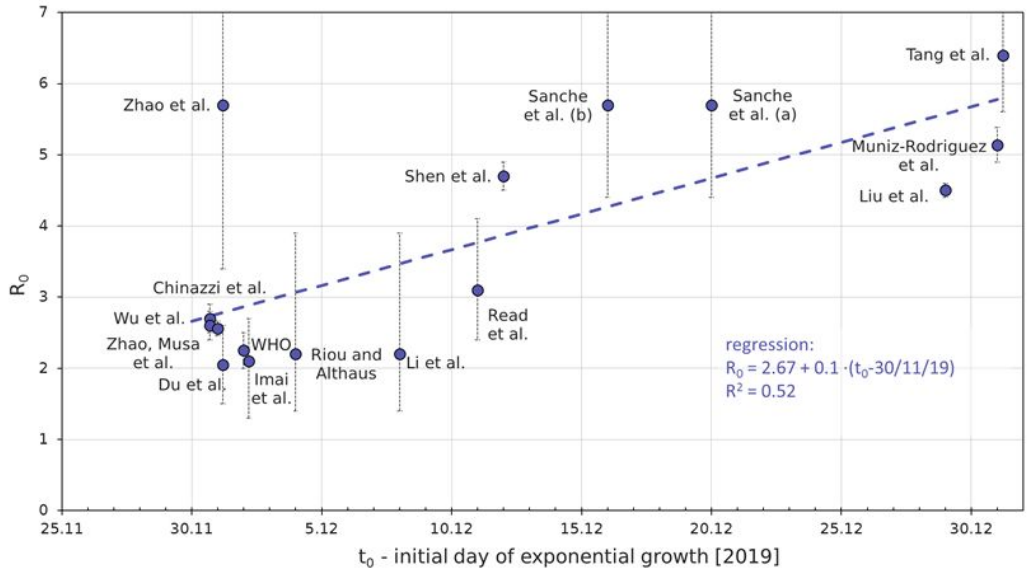

**Figure 2. Correlation between $R_0$ and the assumed initial date of exponential growth.** The X axis gives the initial date of exponential growth assumed in each study. X ticks represent days since 25.11.19. Error bars represent 95% confidence intervals as given in the studies. The connection between $R_0$ and $t_0$ is clear from the linear trend in the scatter plot (least squares linear regression $R^2=0.52$, $P < 10^{-3}$).

### How does isolation of infected patients affect the $R_0$ and generation interval?

Quarantine and contact tracing are common means of slowing epidemics like COVID-19. These strategies involve the isolation of patients known or likely to be infected. When an individual is found to be infected (e.g. by PCR testing), they can be isolated to avoid causing secondary infections. Contact tracers can then identify people who have been in close contact with this patient in recent days so that those contacts can self-isolate until their infectious period has passed. Isolating carriers and their contacts reduces the average number of secondary infections produced from a single infection, bringing this number down from the basic reproduction number ($R_0$, the average number of people an infected person infects with no mitigation) to a lower number termed the effective reproduction number or $R_t$. Isolation also indirectly affects the "generation interval," which is the characteristic time delay between a primary infection and the subsequent transmission event (i.e. infection of a contact). To help conceptualize this effect, consider a case in which an infected person infects 4 others over the course of 4 days, i.e. averaging one person per day over four days. This translates into an $R_t$ of 4 and an average generation interval of 2.5 days. Now let's consider the effect of quarantining infected people 2 days after infection. Now these individuals can only infect others on the first and second days, bringing their personal $R_t$ values to 2, and shortening their generation interval to (1+2)/2 = 1.5 days. Therefore, isolation and contact tracing will tend to shorten the generation interval when compared to earlier stages of the epidemic when no countermeasures were in place (Bi et al. 2020). Thus, isolation and contact tracing will tend to affect the growth of the epidemic both in two ways: reducing secondary transmission and also decreasing the interval between primary and secondary infections known as the generation interval.

When estimating the effectiveness of an intervention, it is common to look at the effective reproduction number ($R_t$), as it can be derived from case counts by various methods. However, most methods for estimating $R_t$ assume that the generation interval stays constant even though there is good evidence against this assumption (Cori et al. 2013; Jacco Wallinga and Teunis 2004). Hence, these approaches to estimating the effective reproduction number don't address the 'physical' meaning of $R_t$: the number of infections an infectious person makes on average, but an 'operational' meaning: the reproduction number derived from the instantaneous growth rate, assuming constant generation interval. Because during the progression of the epidemic the generation interval usually becomes shorter due to interventions, the actual average number of infections an infected person makes (the 'physical' $R_t$) is generally lower than the 'operational' $R_t$ for the same instantaneous growth rate. Operational $R_t$ values may be useful for predicting the trajectory of the epidemic so long as predictions are made via the same model used to infer $R_t$. Still, care should be taken in interpreting the meaning of these non-physical $R_t$ values.

**How is the dispersion of the distribution of secondary infections affected by interventions?**
The distribution of secondary infection counts is also known as the "offspring distribution" and describes how many individuals each infected person infects. The offspring distribution is generally treated as a negative binomial distribution, which is characterized by a mean ($R_t$) and a dispersion parameter k, such that the variance is given by $R_t(1+R_t/k)$. Lower values of k (usually < 1) are associated with a larger number of 'superspreaders' - individuals who infect many more people than the average, $R_t$. Values of k > 1 are associated with a low number of superspreaders, as the offspring distribution is tightly clustered around the mean value $R_t$ when k is large.

Some of the studies looking at the dispersion of the COVID-19 offspring distribution indicate that 'superspreaders' play a large role in COVID-19 transmission. One study found that the dispersion parameter k is < 0.1, indicating that most of the infections (> 80%) are caused by a small fraction of the infected population (by less than 10% of cases) (Endo et al. 2020). Strict interventions like prohibiting large social gatherings, physical distancing, and lockdown mitigate the effect of superpreaders by limiting contact between people. It is of importance to understand how public health interventions affect the dispersion of the offspring distribution. Moreover, since these interventions tend to force people to spend more time at home, studying the offspring distribution before and after interventions like lockdown can teach us about the dynamics of COVID-19 transmission within households, as we discuss below.

Figure 3 shows estimates of the dispersion of the COVID-19 offspring distribution from eight different studies. For each value of k, we also plot the corresponding fraction of infected individuals responsible for 80% of transmissions, calculated assuming a negative binomial distribution. The results are divided according to the degree of public health interventions applied during the study period, i.e. whether strong interventions like lockdown, physical distancing and contact tracing were in place. While strong interventions are in effect the dispersion parameter is in the range of 0.4-1 indicating that 30%-40% of the infected individuals are responsible for 80% of the infections. However, a meta-analysis of data from multiple countries which looked at the spread of COVID-19 prior to intervention gave k << 1 (Endo et al. 2020). The low estimates for k prior to intervention, indicates a much more prominent role of superspreader events, with about 10% of the individuals responsible for 80% of the infections.

This difference between the dispersion prior to interventions and after them is also supported by a study of transmission in Tianjin, China (Y Zhang et al. 2020) finding that government control measures increased the dispersion parameter from 0.14 to 0.77.

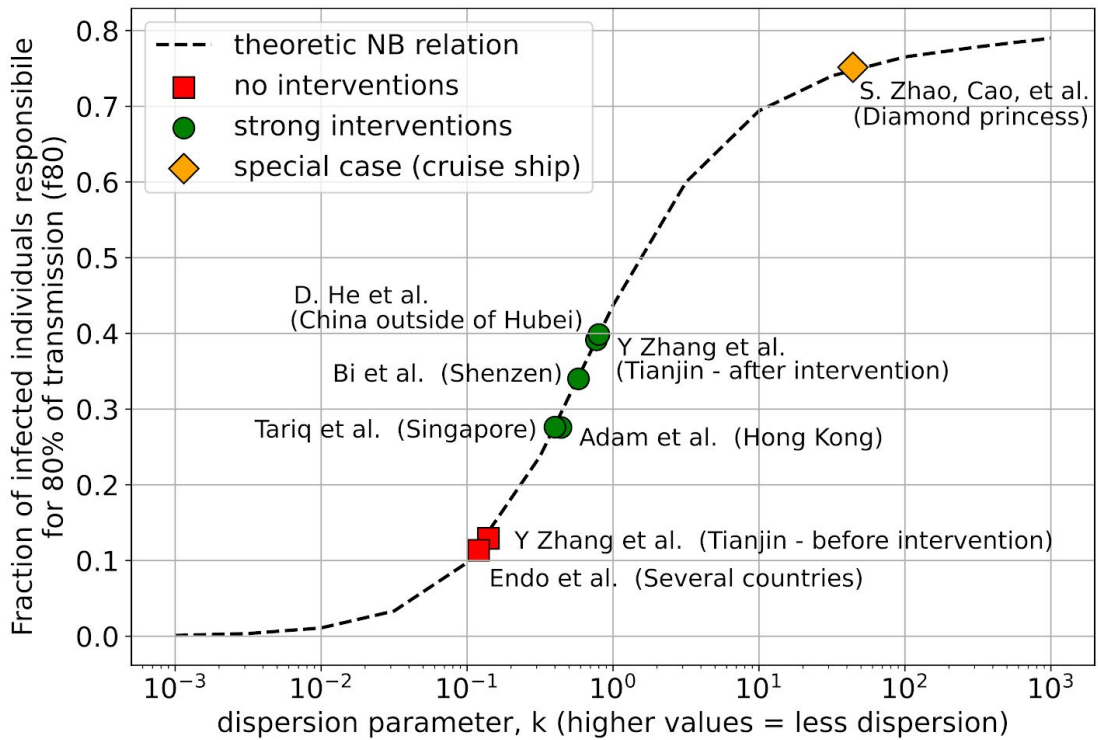

**Figure 3. Estimates of the dispersion parameter of COVID-19 and the corresponding fraction of infected individuals responsible for 80% of transmissions.** The fraction of infected individuals responsible for 80% of transmissions was estimated from a negative binomial distribution following (Lloyd-Smith et al. 2005). The results are divided according to the interventions in the period of the study: cases where physical distancing, contact tracing and lockdowns enacted by governments are indicated in green.

**What can we learn from the household secondary attack rate?**
The secondary attack rate is defined as the fraction of contacts who become infected. The set of people contacted by an infectious individual can be divided into various categories, for example household contacts and external contacts (i.e. outside the household). Making such a distinction is useful for several reasons. First, identifying household contacts is straightforward, which makes determining the secondary attack rate within the household much simpler than determining the overall secondary attack rate. Second, many non-pharmaceutical interventions such as mask usage, physical distancing, or lockdown affect household contacts differently from non household contacts. For example, most people do not wear masks at home. Thus, knowing the household attack rate can provide important data for evaluating the efficiency of non-pharmaceutical interventions.

Let's consider a household with four individuals, one of whom is infectious. Our best estimate for the household secondary attack rate is ≈15% (Bar-On et al. 2020). If we assume most infections in a household are secondary, i.e. a single household member was infected outside the household and now transmits the infection to a totally susceptible household, the infectious household member will infect about 3 x 0.15 ≈ 0.5 others on average. Assuming non-pharmaceutical interventions do not impact the behavior of people inside their homes significantly and that a significant fraction of new infections occur in households with no previous infections, this kind of estimate can be used to estimate a lower bound on the effective reproduction number, $R_t$ ≥ 0.5. Indeed in many places where lockdown was enforced, the effective reproduction number did not decrease much below 0.5 (Flaxman et al. 2020). This calculation also implies that in countries with $R_t$ close to 1, a large portion of infections occur at home (Gudbjartsson et al. 2020). Further measures, such as the isolation of infectious individuals outside their homes, could potentially bring $R_t$ below this figure, as was achieved in several places in China (Leung et al. 2020; Pan et al. 2020). In addition, this calculation makes clear that $R_t$ during lockdown could be significantly influenced by the size of a representative household, which varies by a factor of ≈3 between countries (Bradbury et al. 2014). Because the household reproductive number is dependent on the size of the susceptible population within the household, it would decrease throughout the progression of the epidemic as the pool of susceptible people within a household is depleted. For example, in a household of 4 people, if one person is infected he/she can infect 3 other people. Once one of those people gets infected, they now have only 2 other potential people to infect.

**What are compartmental models and what assumptions do they make?**
Compartmental models are one of the most prominent tools used to study the pandemic and forecast its progression. As the name suggests, these models divide the population into distinct compartments which represent different states of infection. The most commonly-used compartments are: the susceptible compartment (S), containing people who are susceptible to infection but not yet infected; the infectious compartment (I), which includes infected individuals who are contagious; and the recovered compartment (R), which includes people who were infected but are no longer infectious or susceptible, either because they are no longer shedding the virus in high quantities or because they died. These compartments describe the "SIR" model. It is common to supplement these compartments with an additional compartment called "E" for "exposed" individuals, who are not yet infectious. Models with this extra compartment are called SEIR models. Many other elaborations of compartmental models have been generated by the addition of new compartments or parallel compartments treating special subpopulations like essential workers. When considering compartmental models, it is important to note that these models are aimed at describing the transmission of the disease between individuals, and not the clinical progression of the disease within an infected individual. Thus, the infectious period (which is ≈4 days for half-maximum infectivity) describes the duration in which a person is contagious, which is much shorter than the overall duration of illness (on the order of a few weeks), as people's illness is usually prolonged due to the inflammatory response of the immune system. Therefore, people usually stop being strongly contagious well before they recover from the disease.

In SIR and SEIR models, individuals flow through the compartments in a linear fashion - from susceptible (S), to exposed (E), to infected (I), to recovered (R). People are assumed to move between model compartments at a rate proportional to the number of individuals in the compartment. For example, people move from the infectious compartment into the recovered compartment at a constant rate, usually noted as $\gamma$, which is set as the inverse of the average infectious period. In the same manner, people move from the exposed compartment into the infectious compartment at a constant rate denoted $\sigma$, which is set as the inverse of the average latent period. Using such models, however implicitly assumes a specific type of distribution for the latent and infectious periods.

SIR and SEIR models inherently assume that the distribution of latent and infectious periods is exponential, as this is the only distribution for which the probability to transition to the next compartment is constant and independent of the time spent in the current one (known as a "memoryless" distribution). An exponential distribution, however, implies that most people spend a very short time in the exposed or infectious states, which is not consistent with the inference from documented pairs of infectors and infectees that people are typically contagious for ≈4-5 days [(He et al. 2020)](#). Thus, some extended models employ a "trick" to make the distributions more realistic. This trick involves splitting up the exposed and infectious compartments into N sub-compartments each, and multiplying the rate at which people move between those sub-compartments by the N. This way, the sum of people in the exposed sub-compartment represents all the exposed people, and the same holds true for the infectious compartment. Now, however, the time spent in the exposed or infectious compartments will be distributed as a sum of N exponentials, which is called the Erlang distribution. This distribution is similar in shape to the distributions observed when studying infector-infectee pairs [(Ferretti et al. 2020; He et al. 2020)](#).

**How do the testing false positive and false negative rates affect the capacity of organisations to test their constituents and track the disease?**

Given the massive economic implications of lockdown and the low prevalence of COVID-19 in many locations, it is natural for organizations like businesses, government agencies, hospitals and academic institutions to consider comprehensive testing as a means of tracking the disease while resuming partial or full operations. Yet the sensitivity and specificity of current PCR testing methods complicates this approach (see relevant sections on RT-qPCR error rates for numbers and references). Suppose your organization has 1000 employees and you test all of them. If the false positive rate is ≈1%, you are likely to get 10 false positive test results. That is, 10 people who you will presume are infected, but are not. Even if the false positive rate is as low as ≈0.2%, you get 2 false positives on average. False negatives are even more alarming - these represent people who actually have COVID-19 but whose infection you don't detect in a single test. If 0.5% of employees have COVID-19, then there are on average five sick employees. With a 20% false negative rate, you will detect roughly four of five sick employees and miss one who can infect others. In this scenario, you expect to detect 80% of sick individuals (also defined as the 'sensitivity' of the assay), but only 4/14 (≈30%) of your positive test results represent people who are actually infected. Detecting 80% of true infections can be very useful in suppressing the spread, but at the same time it shows one cannot achieve perfect answers with a single test. Rather, the costs and benefits of a testing regime will depend on the error rates and the manner in which positive tests are treated, i.e. whether quarantine and contact tracing are applied. As we discuss below, confirmatory testing by other means (e.g. evaluating symptoms, taking CT scans of the lungs, or group testing approaches) can greatly increase the overall accuracy of a testing regime. Moreover, due to interaction with the wider community, the whole testing operation should be repeated regularly, and in each such iteration one should be cognizant of the implications of false positive and false negative rates.

### How different are your pre- and post-testing odds of having COVID-19?

Because of the relatively high false negative rates associated with PCR testing (10-30%), a negative test does not necessarily imply a clean bill of health. Suppose you live in a location where ≈0.5% of the local population currently has the disease and does not know it. Let's further suppose that you got a negative result from a PCR test. What are the odds that you actually have COVID-19 even though you tested negative? This can be calculated by applying Bayes theorem: $P(C|-) = P(-|C) \times P(C) / P(-)$ where "C" is the event you have COVID-19 and "-" is the event your test was negative. $P(C|-)$ thus denotes the probability that you are a carrier despite a negative test result. The probability of a negative test given that you have COVID-19, $P(-|C)$, is just the false negative rate. Though this value varies widely across literature reports, we will use a 20% false negative rate here (see "RT-qPCR False Negative Rate"). We assumed that the pre-test probability you have the disease, $P(C)$, is 0.5%. The total probability of a negative test, $P(-)$, can be decomposed into contributions from false negatives and true negatives: $P(-) = P(C) \times P(-|C) + P(H) \times P(-|H)$, where "H" is the event that you are healthy. Taking the true negative rate $P(-|H)$ to be 100% minus the false positive rate of ≈1% (0.2-2%, see relevant section) then $P(-) ≈ 0.005 \times 0.2 + 0.995 \times (1-0.01) ≈ 99\%$. We therefore arrive at $P(C|-) ≈ 0.2 \times 0.005 / 0.99 = 0.1\%$. A negative test decreases your odds of having the disease by a factor of roughly 5 (from 0.5% to 0.1%) under this particular set of assumptions, and so a single test cannot fully rule out the possibility of having the virus. This factor of 5 roughly scales as the inverse of the 20% false negative rate we started with (1/0.2 = 5). While this calculation shows the limitations of testing from the perspective of a single person, this calculation of the predictive value of a negative test is also very important from a public health perspective, as we discussed in the previous section.

Let's consider a positive test: what are the odds of actually being COVID-free even though your test was positive? Again, $P(H|+) = P(+|H) \times P(H) / P(+)$ where "H" is the event you are healthy and "+" is the event your test was positive. $P(+|H)$ is just the false positive rate and $P(H)$ was assumed to be ≈ 99.5%. We will assume a false positive rate $P(+|H) ≈ 0.2\%$. This value is at the bottom end of the reported range (0.2-3%, see "RT-qPCR False Positive Rate"), corresponding to the lowest reported value for RT-qPCR testing for influenza [(Merckx et al. 2017)](). We choose this low number because, as of writing, the total positive test rate is around 0.2% in some countries (e.g. Estonia, Denmark, see [Our World In Data for July 3rd 2020]()). Given these values, the total odds of a positive test, $P(+)$, is the sum of true positives and false positives: $P(C) \times P(+|C) + P(H) \times P(+|H) = 0.005 \times (1-0.2) + 0.995 \times 0.002 ≈ 0.6\%$. We can now calculate $P(H|+) = 0.002 \times 0.995 / 0.006 ≈ 33\%$. In words: even when the false positive rate is low (0.2%), the post-test odds of being healthy despite a positive test, $P(H|+)$, can still be quite high. Here we calculated ≈30% odds assuming that 0.5% of the local population is infected, implying that only ≈70% of positive tests came from infected individuals. Moreover, we made an optimistic assumption about the false positive rate. It is important to measure the false positive rate since plausible higher values (e.g. 1%) substantially degrade the information content of positive tests. For example, setting $P(+|H) = 1\%$ gives $P(H|+) ≈ 70\%$, meaning that a positive test would correspond to only 30% odds of having COVID-19.

In practice, PCR testing is not usually applied to individuals chosen at random from the local population. Rather, people often seek testing and treatment when they have COVID-like symptoms or have been in close contact with confirmed cases. Let's assume that there are 20 times more people with COVID-19-like symptoms than symptomatic COVID-19 infectees, i.e. that the COVID-19 infectees are 5% of the symptomatic population (as compared to 0.5% of the population at large). If we recalculate the odds of being COVID-free after a positive test, we get

P(H|+) ≈ 0.002 x (1 - 0.05) / 0.042 ≈ 5%, representing a ≈6-fold decrease in the total number of false positives from ≈30% of all tests to ≈5% in this case. Since testing symptomatic patients is much more informative than testing the population at large, it follows that other means of selecting likely candidates for testing, e.g. by CT scans, other diagnostics, or even a second PCR test, can similarly improve the information content of a COVID-19 testing program.

**What is the effect of double testing patients by RT-qPCR?**
PCR testing of patient samples is the primary mode of diagnosing patients presenting COVID-19 symptoms. In the sections dedicated to the accuracy of PCR testing we gave references supporting an RT-qPCR false negative rate of ≈20% and a false positive rate ≈1%. In general, false negatives are far more concerning than false positives because they are much more likely and also because the consequences are far more serious: a false negative results in improper treatment of an infected individual and, if abundant, false negatives are likely to result in increased transmission. Several sources recommend multiple testing of patient samples to reduce the number of false negatives.

Let's consider a patient who is, in fact, infected with SARS-CoV-2. If repeated RT-qPCR tests are statistically independent and conducted at roughly the same time (relative to the onset of symptoms) then they should each have the same false negative rate of ≈20%. The odds of getting two negative tests given a positive patient should then be $0.2^2$ = 4%. In contrast, (Jiang et al. 2020) report a 22% false negative rate for a single test and 14% for two tests. If tests were independent, the two-test rate should have been reduced to ≈5% (a 17 percentage points reduction) but double-testing gave an empirical false negative rate of ≈14% (an 8 percentage points reduction). So double-testing had roughly half the expected effect in this instance. This calculation implies that PCR tests are not fully independent. One reason that this might be the case is that the second test is always conducted after the first and patient viral loads change over time, typically decreasing below the limit of detection ≈10 days after symptoms (see "Duration of PCR positivity").

Let's consider two subsequent RT-qPCR tests. If we assume that the first test was conducted when the patient had near-maximal viral load (3-4 days after symptom onset, ≈20% false negatives) and the second test was conducted two days later, it might have a substantially greater false negative rate approaching 40-50% as reported in Figure 2 of [(J. Zhao Jr. et al. 2020)](). If the first test has a false negative rate of 22% and the second ≈50%, then the two-test rate should be about 14%, similar to the 14% rate observed. Recent modeling efforts by (Wikramaratna et al. 2020) performed similar, albeit more sophisticated calculations of the effect of multiple testing assuming a 1 day delay between subsequent tests, arriving at false negative rates similar to those empirically observed for two-sample RT-qPCR. As PCR-positivity is observed to decrease over time, their calculation recommends rapid testing with quick turnaround times early in the disease progression (i.e. quickly after symptoms present) to minimize the false negative rate of both single and double tests [(Wikramaratna et al. 2020; Larremore et al. 2020)]().

Another potential explanation for non-indepedence of repeated tests is that there is a physiological cause for false negative results (e.g. certain patients produce less sputum). Physiological differences between patients would also produce correlation between repeated tests.